\documentclass[11pt,twoside]{article}

\usepackage{asp2004}
\usepackage{epsf}
\usepackage{psfig}
\usepackage{lscape}

\begin{document}
\title{Metallicity vs. Be phenomenon relation in the solar neighborhood}
\author{Levenhagen R.S.$^1$, Leister, N.V.$^1$, Zorec, J.$^2$, Fr\'emat, 
Y.$^3$}
\affil{$^1$Instituto de Astronomia, Geof\'{\i}sica e Ci\^encias Atmosf\'ericas 
da Universidade de S\~ao Paulo, Brazil}
\affil{$^2$Institut d'Astrophysique de Paris, UMR7095 CNRS, Univ. P\&M Curie}
\affil{$^3$Royal Observatory of Belgium}

\begin{abstract}
 Fast rotation seems to be the mayor factor to trigger the Be phenomenon. 
Surface fast rotation can be favored by initial formation conditions, such as
abundance of metals. We have observed 118 Be stars up to the apparent 
magnitudes $V=9$ mag. Models of fast rotating atmospheres and evolutionary 
tracks were used to interpret the stellar spectra and to determine the stellar 
fundamental parameters. Since the studied stars are formed in regions that are 
separated enough to imply some non negligible gradient of galactic metallicity, 
we study the effects of possible incidence of this gradient on the nature as 
rotators of the studied stars.
\end{abstract}
\vspace{-0.5cm}

\section{Aim}

 In the present paper we would like to know whether the content of metals in 
the formation regions of stars can introduce some signature in the 
($\tau/\tau_{\rm MS},M/M_{\odot}$) diagram [$\tau/\tau_{\rm MS}\!=$ fractional
age the stars can spent in the main sequence (MS) evolutionary phase]. In fact, 
it was shown by Ste\c pie\'n (2002) that magnetic fields not exceeding 400 G 
can spin up early type stars in the PMS life, while stronger magnetic fields 
spin them down. However, since the mechanism acts through magnetic 
mass-accretion and magnetic-disc locking, the efficacy of the magnetic 
interaction can differ according to the metallic content in the star and its 
gaseous environment.\par
 It has been shown that Be stars rotate at $\Omega/\Omega_c\sim0.9$ (Fr\'emat
et al. 2005). This rotational rate can be attained in the MS evolutionary 
phase only if stars have high rotational rates early in the ZAMS. It is 
expected then that the presence of magnetic fields and its effectiveness at 
establishing high initial stellar surface rotations can be different according 
to the metallic content of the medium where they are formed: {\it age vs. mass}
distributions may then be somewhat different. Our aim is to study Be stars 
situated towards the {\it galactic-center} and in the {\it anti-center}
directions and see whether some information can be drawn form the {\it age vs.
mass} distributions.\par

\section{Method}

 High resolution and S/N spectroscopic observations were carried out on 118 Be 
stars in the Southern Hemisphere with FEROS spectrograph at ESO/La Silla 
(Chile) and with the Coud\'{e} spectrograph at the 1.60m telescope of MCT/LNA 
(Brazil). ESO spectra were taken with a spectral coverage of 3560-9200 \AA\ 
and typical S/N $\sim 200$. LNA spectra were obtained with a WI098 CCD from 
3939~\AA~ to 5060~\AA~ with a reciprocal dispersion of 0.24~\AA/pixel.\par
 The stellar fundamental parameters were obtained in two steps. First, we 
derive the ($T_{\rm eff},\log g$) parameters with classical non-LTE model 
atmospheres that we call $apparent$. In the second step, we correct the
$apparent$ parameters for rotational effects due to stellar deformation and 
the concomitant gravitational darkening effect. We obtain thus the $parent$ 
$non$ $rotating$ $counterpart$ fundamental parameters, i.e. those which depict
homologous stars without rotation (Fr\'emat et al. 2005). We correct them once 
more to obtain the fundamental parameters $averaged$ over the entire deformed
stellar surface, which are finally used to interpolate masses $M$ and ages 
$\tau$ of the studied stars in evolutionary tracks calculated for rotating 
objects (Meynet \& Maeder 2000, Zorec et al. 2005).\par
 A first estimate of the effective temperature and gravity  is attempted 
through the fit of equivalent widths of many spectral lines, including Balmer
lines and line intensity ratios such as He\,{\sc ii}/He\,{\sc i} and Si\,{\sc 
iii}/Si\,{\sc ii}. Once the first guess of ($T_{\rm eff}$, $\log{g}$) is 
established we perform a detailed fit of the observed lines with non-LTE 
spectra synthesized with {\sc tlusty} (Hubeny 1988) and {\sc synspec} (Hubeny, 
Hummer, \& Lanz 1994) codes. Ratios of two neighboring ionization states elude 
sensitivities to abnormal element abundances carried by the fast rotation. 
However, the occurrence of He\,{\sc ii} lines in the visible is constrained to 
stars hotter than B2 ($T_{\rm eff}\!>$ 22000 K). As most B and Be stars in our
sample show only He\,{\sc i} lines, we studied many combinations of He\,{\sc i}
line ratios and used those with the less dependence on the helium abundance. 
Fixing $\log{g}$ values, we calculated synthetic He\,{\sc i} and He\,{\sc ii} 
line profiles for different temperatures from 15000 to 30000 K and helium 
number abundances He/H ratios from 0.001 to 0.3. The analysis performed with 
neutral helium lines (He\,{\sc i}$\lambda\lambda$ 4009, 4026, 4121, 4144, 4388, 
4438, 4471 and 4922) show that only a few combinations of them (He\,{\sc 
i}4922/He\,{\sc i}4026, He\,{\sc i}4438/He\,{\sc i}4144, He\,{\sc 
i}4438/He\,{\sc i}4026, He\,{\sc i}4438/He\,{\sc i}4009, He\,{\sc 
i}4144/He\,{\sc i}4121) are less sensitive to the helium content. The 
$apparent$ projected rotational velocity $V\sin{i}$ was derived from the first 
zero of the Fourier transform of the He\,{\sc i}$\lambda$4471 {\AA} line
profile. The $apparent$ $V\sin{i}$, $T_{\rm eff}$ and $\log{g}$ were then 
corrected for rotational effects (Fr\'emat et al. 2005, Zorec et al. 2005).

\begin{figure}[t]
\centerline{\psfig{file=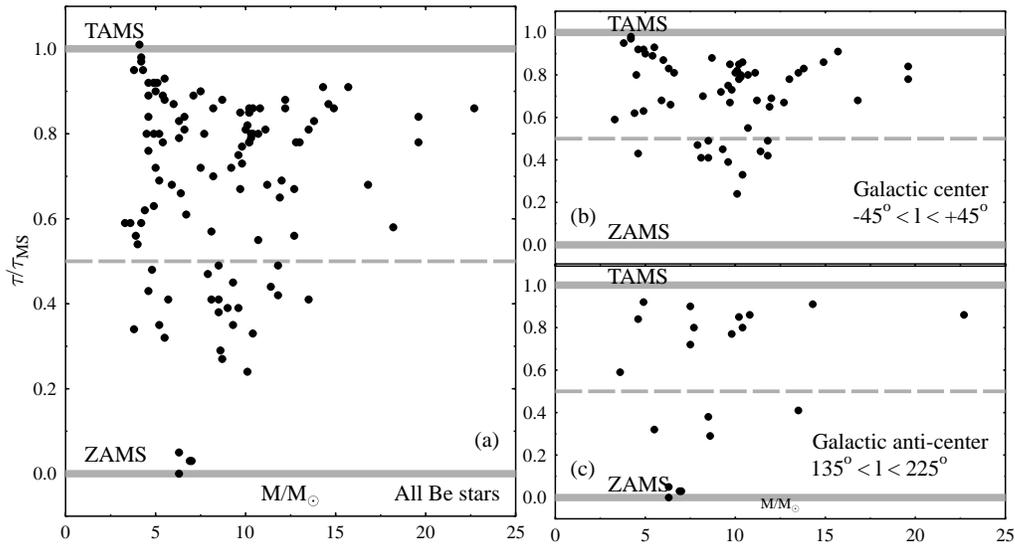,width=13.5truecm,height=7.5truecm}}
\caption[]{(a) Fractional ages $\tau/\tau_{MS}$ ($\tau_{MS}=$ time spent in 
the MS) against the mass of all studied Be stars.(b) Same as (a) for Be stars
located towards the galactic center. (c) Similar to (a) for Be stars located 
towards the galactic anti-center}
\label{f1}
\end{figure}

\section{Results and conclusions}

 The evolutionary tracks used to calculate masses and ages are for rotating 
stars (Meynet \& Maeder 2000, Zorec et al. 2005).  The distribution of 
fractional ages ($\tau/\tau_{\rm MS}\!=$ age/time spent in the MS) against the 
stellar mass of all studied Be stars is shown in Fig.~\ref{f1}a. In this 
figure we see that {\it all studied stars, but one, lay below the TAMS}.
Although most of the studied stars lay in the second half of the MS strip, a
non negligible number of them is still in the first half of the MS evolutionary 
phase. Quite similarly to what was found in Zorec et al. (2005) {\it a lack of 
stars with masses $M \la 7M_{\odot}$ below $\tau/\tau_{\rm MS} =$ 0.5 is 
noticeable}. In our sample, stars with masses $M \ga 12M_{\odot}$ approach the 
TAMS. This may be due to their fast evolution and perhaps to some lack in our
stellar sample of massive Be stars with ages $\tau/\tau_{\rm MS} \la$ 0.5. 
This lack can also be produced by a loss of angular due to their high 
mass-loss rates. They can become low rotators rapidly and thus be impeded to 
display the Be phenomenon any more.\par
 In order to separate the stars in two sets carrying possible information on 
differences in initial matallicity content, we separated them into 
``galactic-center" and ``galactic anti-center" groups. Fig.~\ref{f1}b shows 
that there is no noticeable difference in the {\it ``age vs. mass"} 
distributions thus obtained. The most striking differences are: a) the number
of Be stars in the "galactic-center" group outnumbers the ``anti-center" one;
b) there are younger Be stars in the anti-center direction. Finding a) may 
agree with the fact that low metallicity favors fast rotation (Maeder et al. 
1999). However, the distinction between both groups could be more reliable if 
we could see differences in the number fractions N(Be)/N(B+Be).\par

\end{document}